\documentstyle[12pt]{article}

\begin{document}
\begin{titlepage}
\begin{flushright}
\vbox{EHU-FT-94/5\\
hep-th/9407064}
\end{flushright}
\vskip 5cm
\begin{center}
{\Large\bf Heat kernels and thermodynamics in Rindler space}
\vskip 1cm
R.\ Emparan\\
{\it
Depto. F{\'\i}sica de la Materia Condensada\\
Universidad del Pa{\'\i}s Vasco\\
Apdo.\ 644, 48080 Bilbao, SPAIN
}
\\
{\tt
wmbemgar@lg.ehu.es}
\end{center}
\date{July 1994}
\vskip 1.5cm
\begin{abstract}
We point out that using the heat kernel on a cone to
compute the first quantum correction to the entropy of Rindler
space does not yield the correct temperature dependence.
In order to obtain the physics at arbitrary temperature one
must compute the heat kernel in a geometry with different
topology (without a conical singularity). This is done in
two ways, which are shown to agree with computations
performed by other methods. Also, we discuss the ambiguities
in the regularization procedure.
\end{abstract}
\end{titlepage}
The behavior of quantum fields in black hole backgrounds
has been actively investigated for the last twenty
years \cite{bir}. Thus it may seem surprising that
computations of thermodynamical magnitudes like the entropy
associated to quantum fields in such geometries have been
carried out only recently. It was not until the pioneering
work of 't Hooft \cite{thoo} that the fundamental relevance of
the divergence of these quantities has been fully
appreciated. In recent months there has been a considerable
amount of work analyzing thermodynamical magnitudes in spaces
with horizons. Much of this work has been
carried out in Rindler space, i.e., the space of uniformly
accelerated observers. It shares most of the qualitative
features of black holes and is simple enough to allow detailed
analysis. In particular, the Hawking temperature follows
immediately from the geometry of $D$-dimensional euclidean
Rindler space
\begin{equation}\label{rmetr}
ds^2=\xi^2
d\theta^2+d\xi^2+\sum_{i=1}^N dy_i^2\,.
\end{equation}
Here $y_i$ are the $N=D-2$ transverse flat coordinates,
$\theta$ is the analytically continued euclidean time,
and lines $\xi=$const.\ correspond to uniformly accelerated
observers. The structure of
this manifold is $C_\beta\times M_{D-2}$, where $M_{D-2}$
is the transverse flat space and $C_\beta$ is a flat cone with
angular periodicity $\theta\sim\theta+\beta$. Therefore, to
avoid the conical singularity at the horizon ($\xi=0$),
$\theta$ must have period $\beta_H=2\pi$.

Several methods have been employed to calculate quantum
corrections to the thermodynamical entropy of Rindler space.
In \cite{sus,kab} the statistical mechanics mode counting is
carried out after obtaining the energy spectrum of a scalar
field in Rindler space in the WKB approximation. Since the
density of levels diverges due to the infinite shift of
frequencies near the horizon, a cutoff $\xi_{\rm
min}=\epsilon$ must be introduced. The resulting free energy,
for the massless case in four dimensions, is
\begin{equation}\label{bfree}
{}F(\beta)=-{\pi^2 A\over 180
\epsilon^2\beta^4}\,,
\end{equation}
and shows the characteristic
proportionality to the area. An independent calculation of
Dowker makes use of early determinations of the finite
temperature stress-energy tensor in Rindler space \cite{dow1}
and obtains exactly this same result \cite{dow2}. Temperature
enters here by letting the period $\beta$ take arbitrary
values, and the stress-energy tensor is obtained from the Green
function on $C_\beta\times M_{D-2}$.

The nontrivial topological structure of Rindler space was
exploited by Callan and Wilczek in a variety of ways
\cite{cal}, and allowed them to develop powerful
methods to compute the
entropy. Calculations were performed mostly in two dimensions,
where the divergence is logarithmic and coefficients are
universal (cutoff independent). To go to higher dimensions
they proposed to employ the heat kernel of the Laplacian on
$C_\beta\times M_{D-2}$, defined as
\begin{equation}\label{hkdef} \zeta(\tau)={\rm
tr}\;e^{\tau(\Delta-m^2)}\,,
\end{equation}
from where the free
energy could be obtained:
\begin{equation}\label{hkfree}
{}F(\beta)=-{1\over 2\beta}\int_0^\infty
{d\tau\over\tau}\zeta(\tau)\,.
\end{equation}
Then the
entropy follows from the standard thermodynamical formula
\begin{equation}\label{ent}
S(\beta)=(\beta{\partial\over\partial\beta}-1)\beta F(\beta).
\end{equation}

Now, in a product space the heat
kernel of the Laplacian decomposes into factors corresponding
to each subspace: from $M_{D-2}$ we obtain
$V_{D-2}/(4\pi\tau)^{D-2\over 2}$, whereas on the
cone $C_\beta$ \cite{fur}
\begin{eqnarray}\label{hkcone}
\zeta_{\beta}(\tau)&=&{1\over 4\pi\tau}\biggl\{ {\beta\over
2\pi}\pi R^2+\tau{\beta\over 6} \Bigl(
\bigl({2\pi\over\beta}\bigr)^2-1\Bigr)+\dots\biggr\}
\nonumber\\
&=&({\rm
terms\,linear\,in\,}\beta)+{\pi\over 6\beta}+\dots
\end{eqnarray}
($R$ is a
cutoff for large radius $\xi$, and the dots indicate terms that
vanish for $R\rightarrow\infty$). When computing the free
energy, terms linear in $\beta$ disappear after subtraction of
the $\beta=\infty$ contribution \cite{dow2} (they are
automatically eliminated from the entropy by the prescription
(\ref{ent})). The lower
integration limit in Eq.~(\ref{hkfree}) needs short distance
regularization $\tau>\epsilon^2$.
Then, for general $D>2$ ($N>0$), and $m^2=0$, one
finds
\begin{equation}\label{wfre}
{}F(\beta)\propto {V_N\over
\beta^2\epsilon^N}\,.
\end{equation}

We see that this procedure yields the
correct $V_{N}$ proportionality of the entropy and the
divergence $\epsilon^{-N}$. However, if we compare
Eq.~(\ref{wfre}) (for $N=2$) with Eq.~(\ref{bfree}) we notice
the different dependence on $\beta$. Although the power-like
dependence on the regulator $\epsilon$ precludes direct
comparison of numerical coefficients, the temperature
behavior has a physical significance. The introduction of
temperature by using geometrical concepts may seem obscure,
but the sum over states in statistical mechanics has a clear
physical meaning. In the following we shall construct the heat
kernel that yields thermal magnitudes at arbitrary temperature.

Rindler
space can be regarded as a redefinition of Minkowski space,
but their physical properties differ due to the different
choice of timelike Killing vectors for each space. This leads
to a different definition of positive frequency states,
and therefore, different vacua and Feynman propagators. As
regards to thermodynamics, the Hamiltonians in the partition
function $Z(\beta)={\rm tr}\exp{(-\beta H)}$ are different in
each case. We shall motivate the construction of the heat
kernel starting from the expression of the partition function
as a path integral in euclidean field theory. As
pointed out by Barb\'on in Ref.~\cite{bar}, since the
action is the same in the inertial and accelerated frames,
differences must arise from the choice of the functional
integration measure.

The euclidean action of a massive scalar field
in $D$-dimensional Rindler space can be written as
\begin{equation}\label{act}
S[\phi]=-{1\over 2}\int d\theta\;d^N y\;d\xi\;\xi\;\phi
(\Delta-m^2)\phi\,,
\end{equation}
with $\Delta$ the Laplacian in polar coordinates
(\ref{rmetr}). This way of rewriting
the action of a scalar in flat space singles out the euclidean
time $\theta$, whose periodicity properties will determine the
thermal behavior.

In order to define the appropriate functional
integration for Rindler space we find it convenient to
introduce the ``optical'' metric \cite{gib}
\begin{equation}\label{opmet}
d{\tilde s}^2={ds^2\over g_{00}}=d\theta^2
+\xi^{-2}(d\xi^2+ \sum_{i=1}^N dy_i^2)
\,,
\end{equation}
($\sqrt{\tilde{g}}=\xi^{-N-1}$) and rescale the field by
$\tilde{\phi}=\xi^{N/2}\phi$. The action is now rewritten as
\begin{equation}\label{opact}
S[\tilde{\phi}]= -{1\over 2} \int d\theta \;d^N y
\;d\xi\;\sqrt{\tilde{g}} \;\tilde{\phi} (\widetilde{\Delta}
+{N^2\over 4}-\xi^2
m^2)\tilde{\phi}\,.
\end{equation}
$\widetilde{\Delta}$ is the Laplacian in the optical metric
(\ref{opmet}). An additional
term $N^2/4$ has appeared due to the non-zero curvature
${\tilde R}=N(N+1)$ of the optical space.

Consider now the partition function
over fields periodic under $\theta\sim \theta+ k\beta$
($k\in Z$),
\begin{equation}\label{partf}
Z(\beta) =\int{\cal
D}\tilde{\phi} e^{-S[\tilde{\phi}]} \,,
\end{equation}
with measure ${\cal
D}\tilde{\phi}\sim \prod_{\theta,\xi,y_i}
{\tilde{g}}^{1/4} d\tilde{\phi} (\theta,\xi,y_i) $.
Functional measures are formally specified by some inner
product of functions. Our choice of measure corresponds to the
usual inner product in optical space. This is adequate,
since the spatial section of this product is the same as the
Klein-Gordon product in Rindler space, which, in turn, is
determined by the specific choice of timelike Killing vector.
Actually, we shall see that this will be enough to obtain the
same thermodynamics as the one constructed from Rindler mode
counting \cite{sus,kab}.

Our partition function is thus given
by
\begin{eqnarray}\label{logz}
\log Z(\beta) &=& -{1\over 2}\log{\rm det}
(-\widetilde{\Delta} -{N^2\over 4}+\xi^2 m^2) \nonumber\\
&=& {1\over 2}\int_0^\infty {d\tau\over \tau} {\rm tr}\;
e^{\tau (\widetilde{\Delta} +{N^2/ 4}-\xi^2 m^2)} \,.
\end{eqnarray}
Therefore the object to be calculated is the heat kernel of the
Laplacian in optical space. The optical geometry
is quite different from $C_\beta\times M_{D-2}$. The
structure of (\ref{opmet}) is $S^1\times {\cal H}^{N+1}$, the
hyperbolic space ${\cal H}^{N+1}$ being the $N+1$ dimensional
generalization of the Poincar\'e upper half plane. Euclidean
time has become a coordinate on the circle $S^1$. Thus there
is no conical singularity in this space. Using the
factorization property we see that all the temperature
dependence is contained in the heat kernel of the operator
$(\partial/\partial \theta)^2$ on $S^1$. This is a well known
object. Subtraction of the zero temperature term leaves
\begin{equation}\label{hks1}
\zeta_{S^1}(\tau)={2\beta\over\sqrt{4\pi\tau}}
\sum_{r=1}^\infty e^{-r^2\beta^2/ 4\tau}\,.
\end{equation}

The heat kernel on ${\cal H}^{N+1}$ can be computed using the
DeWitt-Schwinger expansion
\begin{equation}\label{dws}
\zeta(\tau)={1\over(4\pi\tau)^{N+1\over 2}}\sum_{k\geq
0}[a_k] \tau^k \,.
\end{equation}
The coefficients $[a_k]$ can be expressed in terms of volume
integrals of powers and derivatives of the curvature tensor.
In particular, $[a_0]$ is given by the volume of ${\cal
H}^{N+1}$. This is a divergent quantity and needs to be
regularized: we restrict the transverse $y_i$ space to a box
of volume $V_N$ and introduce a lower bound $\epsilon$ for
$\xi$:
\begin{equation}\label{vol}
[a_0]=V_{{\cal H}^{N+1}}={V_N\over
N\epsilon^N}\,.
\end{equation}
(for $N=0$ we find the familiar logarithmic divergence).

Higher coefficients are simpler in the massless four
dimensional case. In this case
$[a_1]$ vanishes, whereas the contribution of $[a_2]$ to
$F(\beta)$ is a constant independent of the temperature, which
can be subtracted out. Regarding higher coefficients, their
contribution to $F(\beta)$ vanishes when zeta-function
regularization of sums is employed. Therefore, we will discard
all these terms. Below we shall find by an independent method
that these terms do not appear indeed. Thus, the heat kernel
appropriate for Rindler space thermodynamics is
\begin{equation}\label{hkd4}
\zeta_R(\tau)={A\beta\over \epsilon^2}{1\over (4\pi\tau)^2}
\sum_{r\geq 1}e^{-r^2\beta^2/ 4\tau}\,.
\end{equation}
Substitute now Eq.~(\ref{hkd4}) into Eq.~(\ref{hkfree}). The
temperature dependence can be easily extracted out by
rescaling $\tau\rightarrow \beta^2\tau$. The remaining
integral and series yield a numerical factor that is easy to
compute. Eq.~(\ref{bfree}) is then exactly reproduced.

We see that in this
approach the origin of divergences is traced to the infinite
volume of optical space. This conclusion was also arrived at by
different means in \cite{bar}.

We now rederive this
result following a different approach, which
emphasizes other quantum aspects
of Rindler space.

Take a basis of eigenfunctions of $-\widetilde{\Delta}
-N^2/4 +\xi^2 m^2$:
\begin{equation}\label{eigf}
\psi_{np\nu}(x)=
{\sqrt{2\nu\sinh\pi\nu}\over (2\pi)^{N+1\over 2}\pi} e^{i2\pi
n\theta/\beta +ip_i y_i}\xi^{N/2} K_{i\nu}(\mu\xi)\,.
\end{equation}
We have denoted $x=(\theta,\xi,y_i)$ and
$\mu=\sqrt{p^2+m^2}$; $K_{i\nu}$ is the modified Bessel
function. The corresponding eigenvalues are
($4\pi^2n^2/\beta^2+\nu^2$). The functions (\ref{eigf}) are
orthonormal with respect to the volume $D$-form of the optical
geometry. (The partition function (\ref{partf}) and its
integration measure become especially simple when
$\tilde{\phi}$ is expanded in this basis).

As is known, all the relevant information can be encoded in
the heat function
\begin{equation}\label{hfun}
K(x',x'';\tau)=\sum_{np\nu}e^{-\tau (4\pi^2 n^2/\beta^2
+\nu^2)} \psi_{np\nu}^\ast (x') \psi_{np\nu}(x'')\,.
\end{equation}
For example, the euclidean two point Green function in
Rindler space can be obtained as
\begin{equation}\label{gree}
G_E(x',x'')=(\xi'\xi'')^{-N/2}\int_0^\infty d\tau\;
K(x',x'';\tau)\,,
\end{equation}
which can be readily seen to agree with previous calculations
in the literature.

The integrated
heat kernel follows from tracing over spacetime indices
\begin{equation}\label{hker}
\zeta(\tau)= \int d^D x \sqrt{\tilde{g}} K(x,x;\tau)\,.
\end{equation}
Now it is interesting to see how divergences appear in this
approach. The eigenvalue problem in
Rindler space is not well defined unless we introduce a cutoff
near the horizon. Aside from discretizing the spectrum, this
has the effect that the  spectrum of frequencies $\nu$ becomes
dependent on the transverse momentum $p$ and bounded below by
a non zero value, $\nu\geq \epsilon\mu$. Also, inner products
of eigenfunctions are now determined by
\begin{equation}\label{norm}
\int_\epsilon^\infty {d\xi\over \xi}\; K_{i\nu}(\mu\xi)
K_{i\nu}(\mu\xi) = {\pi^2\over 2 \nu \sinh \pi\nu}{dn\over
d\nu}\,.
\end{equation}
The eigenvalue density
$dn/d\nu$ has been computed in \cite{sus} using the WKB
method, and is seen to
diverge for $\epsilon\rightarrow 0$. Substitution leads,
after Poisson resummation, to the same heat kernel found above.

Some remarks are now in order. Notice that the regularization
procedure employed when obtaining Eq.~(\ref{wfre}) ---i.e.,
cutting off short proper times $\tau$--- is different from the
prescription $\xi_{\rm min}=\epsilon$ in
Eqs.~(\ref{bfree}) and (\ref{hkd4}). In fact, calculations
using the Rindler metric (\ref{rmetr}) admit both
regularizations \cite{kcom}, but there does not seem to be
any {\it a priori} reason to choose one of them.
On the other hand, when computations are performed in the
optical geometry the regularization procedure is automatically
selected, in much the same way as any infinite volume
regularization. Also, the optical geometry allows for a
physical interpretation in terms of a sum over states through
the factor $\zeta_{S^1}(\tau)$, Eq.~(\ref{hks1}).

One last remark. Rindler space can be regarded
as the large mass limit of the Schwarzschild black hole. For a
finite mass black hole with euclidean geometry
$R^2\times S^{D-2}$, the optical rescaling leads to a
spacetime with a $S^1$ factor. This yields the temperature
dependence through the same function as
Eq.~(\ref{hks1}). Computation of the remaining factor
becomes more involved, but the qualitative features are not
different. Actually, the same arguments can be applied to
find that only $[a_0]$ and $[a_1]$ are needed (in four
dimensions). Also, the divergences of optical volume that
appear in $[a_0]$ are seen to be general.

\bigskip
I would like to thank M.\ A.\ Go\~ ni, J.\ L.\ Ma\~ nes and
M.\ A.\ Valle for useful discussions and encouragement. This
work has been partially supported by a FPI grant from MEC
(Spain) and projects UPV 063.320-EB119-92 and CICYT
AEN93-0435.

\end{document}